\documentclass[]{quantumarticle}

\usepackage{amsmath,amssymb}
\usepackage[]{graphicx}
\usepackage[utf8]{inputenc}
\usepackage[colorlinks,linkcolor=blue,citecolor=blue,urlcolor=blue,breaklinks=true]{hyperref}
\usepackage{cleveref}
\usepackage{soul}
\usepackage{color}
\usepackage{xcolor}
\usepackage{tikz}
\usetikzlibrary{shapes,tikzmark}

\usepackage[textsize=tiny]{todonotes}

 \newcommand{\bra}[1]{\left\langle #1 \right |}
 \newcommand{\ket}[1]{\left | #1 \right\rangle}

\graphicspath{{./figures/}}

\definecolor{1sol}{HTML}{0d0887}
\definecolor{3sol}{HTML}{9c179e}
\definecolor{5sol}{HTML}{ed7953}
\definecolor{7sol}{HTML}{f0f921}

\definecolor{blue}{HTML}{1f77b4}
\definecolor{orange}{HTML}{ff7f0e}
\definecolor{green}{HTML}{2ca02c}
\definecolor{red}{HTML}{d62728}
\definecolor{purple}{HTML}{9467bd}
\definecolor{brown}{HTML}{8c564b}
\definecolor{pink}{HTML}{e377c2}
\definecolor{grey}{HTML}{7f7f7f}
\definecolor{khaki}{HTML}{bcbd22}
\definecolor{cyan}{HTML}{17becf}

\begin{document}

\title[Ground-state bistability]{Quantum bistability in the hyperfine ground state of atoms}
\author{B. G\'abor}
\address{Wigner Research Centre for Physics, H-1525 Budapest, P.O. Box 49., Hungary}
\author{D. Nagy}
\address{Wigner Research Centre for Physics, H-1525 Budapest, P.O. Box 49., Hungary}
\author{A. Vukics}
\address{Wigner Research Centre for Physics, H-1525 Budapest, P.O. Box 49., Hungary}
\orcid{0000-0001-8916-4033}
\email{vukics.andras@wigner.hu}
\author{P. Domokos}
\address{Wigner Research Centre for Physics, H-1525 Budapest, P.O. Box 49., Hungary}
\orcid{0000-0002-1002-5733}


\begin{abstract}
First order phase transitions are ubiquitous in nature, however, this notion is ambiguous and highly debated in the case of quantum systems out of thermal equilibrium. We construct a paradigmatic example which allows for elucidating the key concepts.  We show that atoms in an optical cavity can manifest a first-order dissipative phase transition where the stable co-existing phases are quantum states with high quantum purity. These states include hyperfine ground states of atoms and coherent states of electromagnetic field modes. The scheme benefits from the collective enhancement of the coupling between the atoms and the cavity field. Thereby we propose a readily feasible experimental scheme to study the dissipative phase transition phenomenology in the quantum limit, allowing for, in particular, performing a finite-size scaling to the  thermodynamic limit.
\end{abstract}


\maketitle

First-order dissipative quantum phase transitions (DQPT) \cite{Kessler2012,Macieszczak2016} feature the following defining properties: (i) in a finite range of a given control parameter, the quantum system has multiple stable steady-states, (ii) which are macroscopically discernible by an order parameter, and (iii) are approximately pure quantum states. When sweeping the control parameter across the critical domain, the steady state depends on the history and the order parameter exhibits a hysteresis. It is condition (iii) which is exotic, and while many classical systems exhibit the multistability conditions (i) and (ii), it is only recently that examples for the quantum version have been found in various systems. A first-order DQPT was predicted theoretically for the clustering of Rydberg-atoms \cite{Lee2012,Ates2012,Marcuzzi2014}, although the experimental feasibility has been contested \cite{Carr2013,Malossi2014,Urvoy2015,Letscher2017}. Optical lattices with engineered losses \cite{Labouvie2016,Benary_2022}, ultracold-atom cavity QED systems \cite{Ferri2021}, nonlinear photonic or polaritonic modes \cite{rodriguez_probing_2017,fink_signatures_2018}, exciton-polariton condensates \cite{Dagvadorj2021}, and circuit QED systems \cite{fink_observation_2017,Fitzpatrick2017,Sett2022} have also been shown to feature first-order dissipative phase transitions.

Externally driven dissipative quantum systems can evolve to steady states in a balance of the driving and dissipation, which is different from a thermodynamic equilibrium. In these steady states there is a steady current, e.g. photon flux through the system which provides for a measurable macroscopic signature of the steady states. Measurement is a substantial component for using the phase transition terminology. A typical experimental setting for this phenomenology is cavity quantum electrodynamics (QED). In the optical domain of cavity QED, atoms are loaded into a Fabry--P\'erot-type optical resonator which is driven by a laser source. The light partially transmitted through the mirrors carries information about the intra-cavity field created in an interaction with the atoms. Continuous  monitoring of the outcoupled light by a photodetector results in a measurement signal of the intensity, which serves as an order parameter of the intra-cavity system. Even if the system has only a few degrees of freedom, i.e., it is not macroscopic, the steady states can be considered phases because of the macroscopic measurement signal. 

Optical bistability \cite{Bonifacio1978photon,Bonifacio1978bistable,dombi_optical_2013,Casteels2017,Heugel2019} is a paradigmatic example of a first-order phase transition in cavity QED.  In its common form \cite{Bonifacio1978bistable} it satisfies the first two conditions, (i) and (ii), but not the third one (iii). The transmission of a laser-driven optical resonator mode can be suppressed or allowed at the same drive intensity, depending on the state of the atoms loaded into the mode volume. The nonlinearity originates from the saturation effect of two-level atoms. When fully saturated, the atomic medium is transparent for the light, which can then fill the resonantly driven cavity mode in a self-consistent manner. A  very different solution can co-exist in a given range of the drive intensity, namely, that the atoms are  close to their ground state and amount to a very absorptive medium.  Light intensity is then suppressed securing the low excitation of the atoms. In this phase the state of the atoms is close to a pure quantum state, i.e., the ground state. However, in the bright phase the saturated atoms correspond to a high-entropy mixed state, breaking condition (iii) above. 

All the three conditions are met in the  case of the photon-blockade breakdown (PBB) phase transition \cite{dombi_bistability_2015,carmichael_breakdown_2015,vukics_finite_2019}. The experimental configuration is very similar, however,  the atomic medium comprises only a single two-level atom and its electric dipole coupling with the mode has to be very large. This can be achieved in superconducting circuit QED systems in the microwave frequency regime where the signatures of a closely related bistability effect have been observed with three-level atoms \cite{fink_observation_2017}. 

In this paper we present an optical cavity QED scheme which manifests a first-order dissipative quantum phase transition in an intrinsic form and makes use of the collective enhancement of the coupling between the mode and an ensemble of atoms. Here, reaching the critical domain also necessitates large cooperativity, but in contrast to PBB, this can be achieved not only by increasing the (single-atom) coupling constant – a feat that appears impossible in the optical domain – but the atom number.  However, the increase of the atom number is not accompanied by an increase of the volume  of the atom cloud,  spatial dimensions do not play an inherent role. 

The proposed scheme is the extension of a recent experiment in which we observed transmission blockade breakdown \cite{Clark2022,Gabor2022}, cf. also \cite{Suarez2023collective} for a related scenario. The extension consists of the use of two cavity modes instead of just one as in the experiment \cite{Gabor2022}. The modes interact with an ensemble of atoms modelled by a four-level scheme. The phases correspond to steady states with high quantum purity. The two modes are either in vacuum or in a high-intensity coherent state, meanwhile the atoms are in one of their hyperfine ground states. Because of this simplicity of the steady states, a semiclassical theory suffices to describe the phase diagram and to calculate the order parameter as a function of the control parameters of the system.
 
\begin{figure}[htb]
\parbox[c]{\columnwidth}{
\includegraphics[width=\columnwidth]{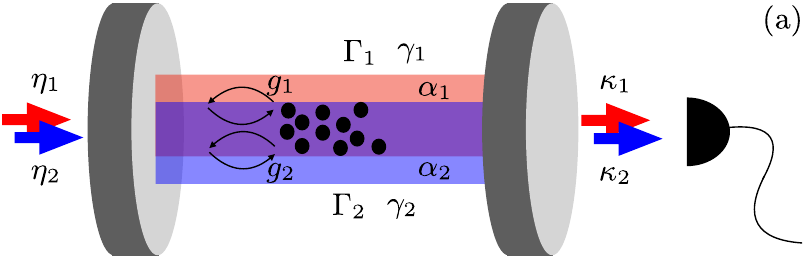}}
\parbox[c]{\columnwidth}{
\includegraphics[width=\columnwidth]{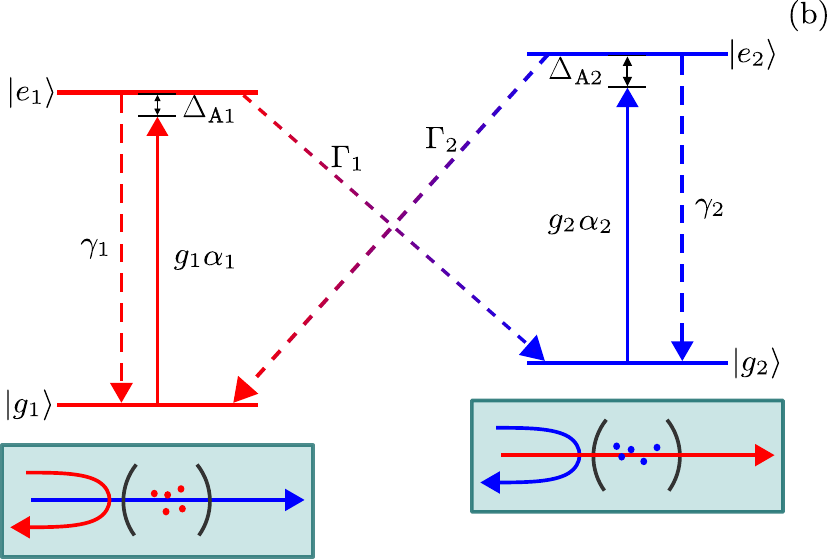}}
\caption{(a) Parameters of the cavity QED scheme with two laser-driven modes interacting with an ensemble of atoms. The modes are spatially separated only for illustrative purposes, in practice, two fundamental modes of the cavity can be used. (b) Relevant level scheme of the atoms with two dipole-allowed transitions cross-coupled by relaxation processes. Schematic panels at the bottom represent that the atomic ground states switch  the  transmission or reflection of the cavity drives.}
\end{figure}

Consider a cold ensemble of $N$ atoms interacting with two modes of a cavity with frequencies $\omega_{\mathrm{C}i}$ and linewidths $\kappa_i$ ($i=1,2$ for the two modes, respectively). The modes are externally driven by coherent laser lights at frequencies $\omega_i$,  with driving amplitudes $\sqrt{N}\,\eta_i$, respectively. The scaling factor $\sqrt{N}$ is introduced for later convenience.  The cavity modes couple to the electric dipole transitions $\ket{g_1}\leftrightarrow\ket{e_1}$ and $\ket{g_2}\leftrightarrow\ket{e_2}$, respectively. These transitions have resonance frequencies $\omega_{\mathrm{A}i}$, linewidths $\gamma_{i}$,  and the electric dipole interaction strength is expressed in terms of the single-photon Rabi frequency, $g_i=\sqrt{\tfrac{N\,\omega_{\mathrm{C}i}}{2\epsilon_0\hbar {\cal V}}} d_{i}$, with $d_{i}$ being the atomic dipole moment, and the mode volume is ${\cal V}$. Note that by introducing $N$ into the definition of the $g_i$, these parameters describe the collective coupling between the modes and the cold atomic ensemble.

Importantly for the scheme, atoms from the excited levels may decay not only to the cavity-coupled ground state, i.e. $\ket{e_i}\rightarrow\ket{g_i}$, but cross decays $\ket{e_1}\rightarrow\ket{g_2}$, $\ket{e_2}\rightarrow\ket{g_1}$ are also possible with rates $\Gamma_1$ and  $\Gamma_2$, respectively. This is the mechanism that couples the subspaces 1 and 2.


The time evolution of the dynamical variables can be described by a semiclassical mean-field model using Maxwell-Bloch equations.  Let $\alpha_i=\text{Tr}(a_i)/\sqrt{N}$ denote the complex amplitude of the cavity field modes, where $a_i$ is the annihilation operator of the respective cavity mode.   The atomic polarization of the corresponding transition is the quantum average $m_i=\left\langle\ket{g_i}\bra{e_i}\right\rangle$. In the spirit of the mean-field approximation, the atoms are not distinguished according to their coordinates or velocities in the cavity mode, so we do not need to add ensemble averaging to the collective variables. The filling ratios of the states $\ket{g_i}$ are denoted by $n_{\mathrm{g}i}=\left\langle\ket{g_i}\bra{g_i}\right\rangle$ – they are numbers between 0 and 1 – and $n_{\mathrm{e}i}$ for the states $\ket{e_i}$ are defined analogously. The mean-field equations of motion read
\begin{subequations}
\label{eq:meanfield}
\begin{align}
\label{eq:meanfield1}
  \dot \alpha_1 &=  (i \Delta_{\mathrm{C}1}-\kappa_1 ) \alpha_1 + g_1\,m_1 + \eta_1\,,\nonumber\\
 \dot m_1&= (i \Delta_{\mathrm{A}1} -\gamma_1 -\Gamma_1 )\,m_1 + g_1\left[n_{\mathrm{e}1}-n_{\mathrm{g}1}\right]\alpha_1\,, \nonumber\\
  \dot n_{\mathrm{e}1}&= - g_1\left[\alpha_1^* m_1+ m_1^* \alpha_1\right]-
            2 (\gamma_1 + \Gamma_1)  n_{\mathrm{e}1}\,,\nonumber\\
  \dot n_{\mathrm{g}1}&=g_1\left[\alpha_1^* m_1+m_1^* \alpha_1 \right]+
          2 \gamma_1\,n_{\mathrm{e}1}  + 2\Gamma_2\,n_{\mathrm{e}2} \,,
\end{align}
\begin{align}
\label{eq:meanfield2}
  \dot \alpha_2 &=  (i \Delta_{\mathrm{C}2}-\kappa_2 ) \alpha_2 + g_2m_2 + \eta_2\,,\nonumber\\
 \dot m_2&= (i \Delta_{\mathrm{A}2} -\gamma_2 -\Gamma_2 )\,m_2 + g_2\left[n_{\mathrm{e}2}-n_{\mathrm{g}2}\right]\alpha_2\,, \nonumber\\
 \dot n_{\mathrm{e}2}&= - g_2\left[\alpha_2^* m_2+ m_2^* \alpha_2\right]-
            2 (\gamma_2 + \Gamma_2)  n_{\mathrm{e}2}\,,\nonumber\\
  \dot n_{\mathrm{g}2}&=g_2\left[\alpha_2^* m_2+m_2^* \alpha_2 \right]+
          2 \gamma_2\,n_{\mathrm{e}2}  + 2\Gamma_1\,n_{\mathrm{e}1}.
\end{align}
\end{subequations}
The equations of the field mode $\alpha_i$ and the polarization $m_i$ are written in a frame rotating at $\omega_i$ ($i=1,2$). 
Without loss of generality, for simplicity, we will consider a symmetric case that the parameters with index $i=1$ and $2$ are equal pairwise, ${\gamma_i=\gamma}$, ${\Gamma_i=\Gamma}$, ${\kappa_i=\kappa}$, and ${g_i=g}$ for $i=1,2$.  We consider resonant driving of the cavity modes, ${\Delta_{\mathrm{C}1}=\Delta_{\mathrm{C}2}=0}$. The cavity linewidth $\kappa=1.32\gamma$ and the atom-cavity coupling are taken from the experiment \cite{Gabor2022}, for this latter the single atom coupling $g(N=1)=0.1 \gamma$. Without loss of generality we chose $\Gamma=\gamma$. The drive amplitudes $\eta_1$ and $\eta_2$ are left to be the control parameters of the system which can be tuned to explore different phases and transitions between them.

What we achieved with the scaling with $N$ of the dynamical variables and parameters introduced above is that $N$ does not appear in the system (\ref{eq:meanfield}), not even as the upper limit of the range of the population variables. Moreover, $g$ with the above definition (incorporating a factor of $\sqrt{N}$) makes $g^2$ proportional to the ensemble cooperativity ${\cal C} \equiv g^2/\sqrt{(\Delta_\mathrm{C}^2+\kappa^2)(\Delta_\mathrm{A}^2 +\gamma^2)}$. This latter quantity is a measure of nonlinearity as attested by that optical bistability in a system of \emph{two-level atoms} coupled to a cavity mode becomes possible in the ${\cal C} \sim 1$ regime. Note that for $\Gamma_i=0$ the system (\ref{eq:meanfield}) separates to two uncoupled two-level systems, where bistability would originate from the saturation of the atoms. Although we will consider large atomic detuning $\Delta_\mathrm{A}$ with respect to the linewidth $\gamma$ (for numerical calculations $\Delta_\mathrm{A}=-12\gamma$ was chosen, where the negative sign stands for red detuning), significant excited state population $n_{\mathrm{e}i}$ can occur for large intensities. This possibility is taken into account in these equations. Nevertheless, in the following we will study another solution of the mean-field equations which is bound to the cross-coupling decay terms, and takes place in the low-excitation limit of the atoms.

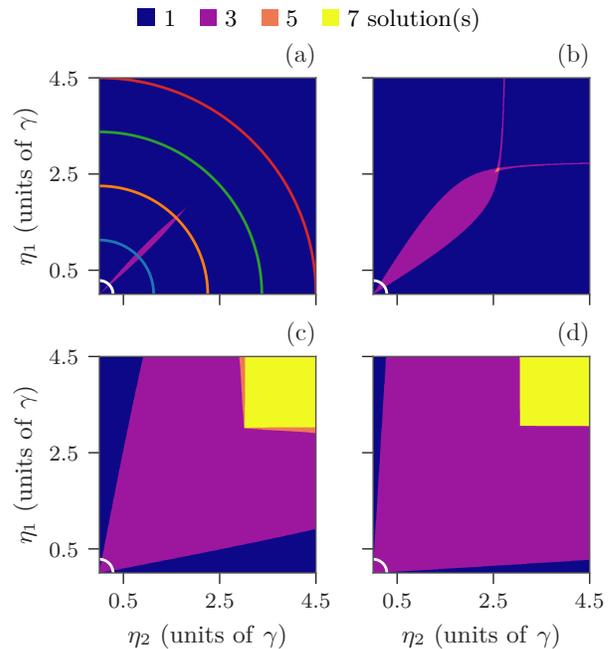
\begin{figure}[htb]
\raisebox{0pt}{\tikz{\node[scale=1,rectangle,fill=1sol](){};}} \small1\quad
\raisebox{0pt}{\tikz{\node[scale=1,rectangle,fill=3sol](){};}}  \small3\quad
\raisebox{0pt}{\tikz{\node[scale=1,rectangle,fill=5sol](){};}}  \small5\quad
\raisebox{0pt}{\tikz{\node[scale=1,rectangle,fill=7sol](){};}}  \small7 solution(s)
\centering
\includegraphics[width=0.98\columnwidth]{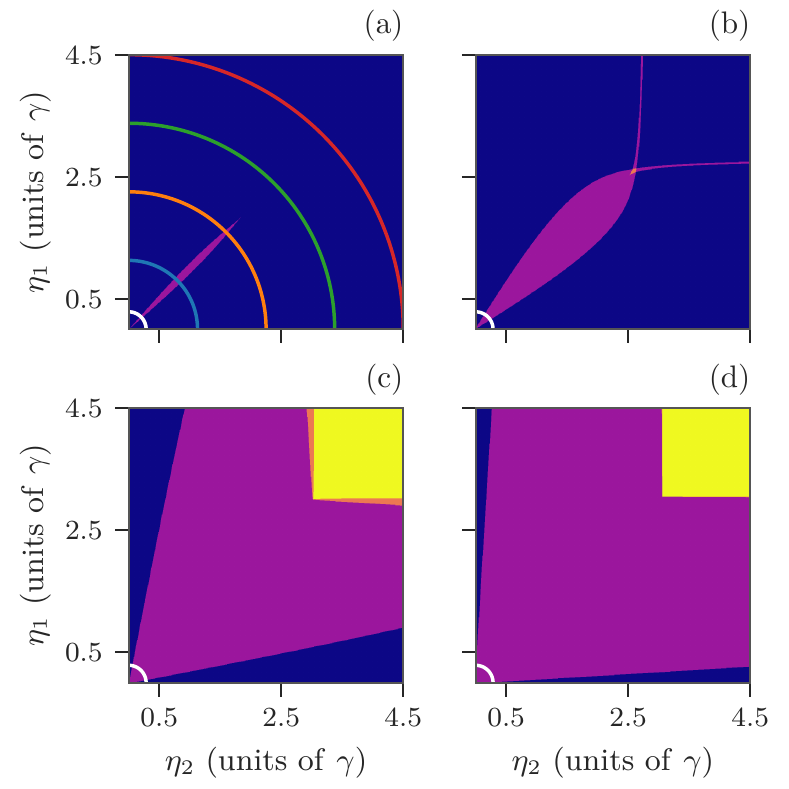}
\caption{Phase diagram with domains with different number of stable solutions of the system (\ref{eq:meanfield}) on the plane of the drive amplitudes $\eta_1$, $\eta_2$. The cooperativity increases from (a) to (d), corresponding to atom numbers $N=5\cdot10^3$, $10^4$, $10^5$, and $10^6$, respectively, with the single atom coupling $g(N=1)=0.1 \gamma$.  Relevant quantities along the coloured quarter circular arcs plotted on the phase diagrams will be shown later in Figs.~\ref{fig:n5000} and \ref{fig:nall}. Radii of the coloured circular arcs are $\eta/\gamma=0.29$, ${1.13}$, ${2.25}$, ${3.38}$, ${4.5}$. The white arcs in the low drive limit are of particular interest with respect to DQPT.}
\label{fig:solnums}
\end{figure}

The steady state solution of Eqs.~(\ref{eq:meanfield}) can be obtained by setting the temporal derivatives on the left hand side to zero. The remaining system of algebraic equations can be transformed into a single, $7$th order polynomial equation with real coefficients for the variable ${n_{\mathrm{e}1}-n_{\mathrm{g}1}}$. Such an equation can have 1, 3, 5, or 7 real solutions out of which respectively 1, 2, 3, or 4 are stable, the rest are unstable. The number of stable solutions depends on the control parameters $\eta_1$ and $\eta_2$, and domains with different numbers are depicted as a phase diagram in Figure \ref{fig:solnums}.

The phase diagram depends on the cooperativity that in the present setup can be changed by the atom number $N$. Whereas $N=5\cdot10^3$ (panel (a)) allows for 1 or 3 solutions only, the large atom numbers $N=10^5$ and $10^6$ (panels (c) and (d)) give rise to domains with 5 (orange edge of the bright yellow domain) or even 7 solutions (bright yellow domain). An intermediate phase diagram is obtained for $N=10^4$ (panel (b)) where a domain with 5 solutions exist, but one with 7 solutions does not. Closer look at the concrete solutions in the domains with 5 and 7 solutions  (not shown here) reveal  that the excited states $n_{\mathrm{e}i}$ are significantly populated, while the polarizations $m_i$ have low values.  This means that the steady states correspond to statistical mixtures, i.e., the quantum purity of the state is low. In the following we will focus on the bottom left corner of the phase diagrams where only one or two stable solutions exist. 

\begin{figure*}
\raisebox{3pt}{\tikz{\node[scale=1,rectangle, inner sep=1 pt,minimum height=0.5pt,minimum width=9 pt,fill=blue](){};}} $\eta/\gamma=1.13$ \quad
\raisebox{3pt}{\tikz{\node[scale=1,rectangle, inner sep=1 pt,minimum height=0.5pt,minimum width=9 pt,fill=orange](){};}} $\eta/\gamma=2.25$\quad
\raisebox{3pt}{\tikz{\node[scale=1,rectangle, inner sep=1 pt,minimum height=0.5pt,minimum width=9 pt,fill=green](){};}} $\eta/\gamma=3.38$\quad
\raisebox{3pt}{\tikz{\node[scale=1,rectangle, inner sep=1 pt,minimum height=0.5pt,minimum width=9 pt,fill=red](){};}} $\eta/\gamma=4.5$
\centering

\includegraphics[width=0.3\linewidth]{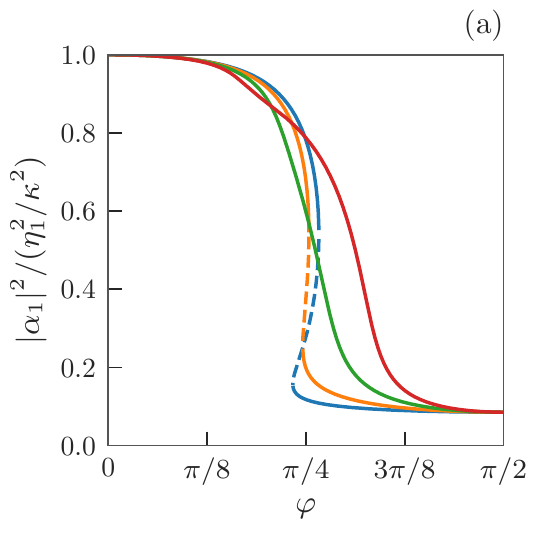}
\includegraphics[width=0.3\linewidth]{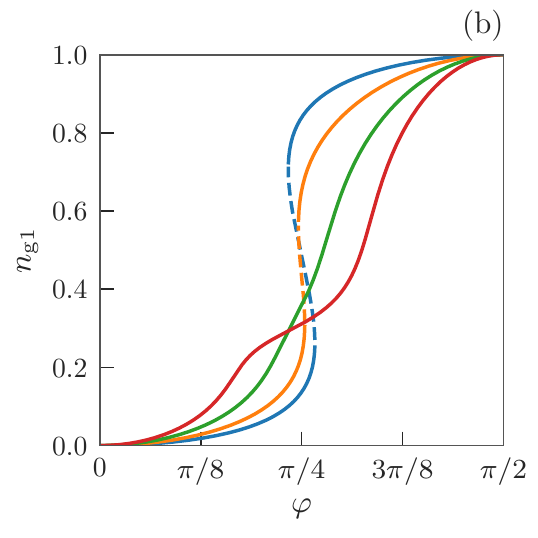}
\includegraphics[width=0.3\linewidth]{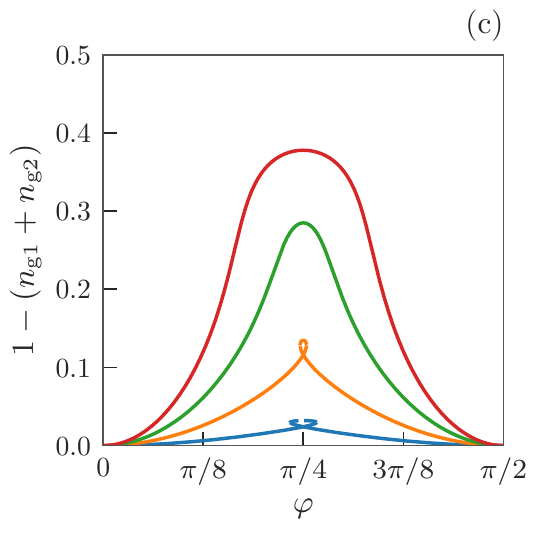}
\caption{Crossing domains with multiple stable solutions. The transmittance of cavity for mode 1 (a), and relative atomic population (b-c) are shown along the circular arcs of corresponding colours plotted in Figure \ref{fig:solnums} (a). Arc measure $\varphi$ is measured from the vertical axis. Solid (dashed) lines correspond to stable (unstable) solutions.}
\label{fig:n5000}
\end{figure*}

The different solutions in a given domain of the phase diagram are distinct in a macroscopic observable, which is the transmitted power $\kappa |\alpha_i|^2$ in our case ($i=1,2$). This is a suitable order parameter of phases and is readily obtained from the mean-field model. The solution (valid for our case of $\Delta_C=0$ and $|\Delta_\mathrm{A}|\gg\gamma + \Gamma$) reads
\begin{equation}
\label{eq:pow}
|\alpha_i|^2=\frac{\eta_i^2}{\kappa^2}\, \frac{1}{1+{\cal C}^2(n_{\mathrm{e}i}-n_{\mathrm{g}i})^2},
\end{equation}
highlighting the role of the cooperativity as a measure of nonlinearity. The factor $\eta_i^2/\kappa^2$ is simply the number of photons in the resonantly driven empty  cavity and will be used as a normalization factor. The second factor above can be identified as transmittance. Equation~(\ref{eq:pow}) is not an explicit solution as the population difference $n_{\mathrm{e}i}-n_{\mathrm{g}i}$ depends on the intracavity intensity $|\alpha_i|^2$. However, this form allows for getting insight to the phases.

If the population $n_{\mathrm{g}1} \simeq 1 $ and $n_{\mathrm{e}1} \ll n_{\mathrm{g}1}$, the transmittance through the mode $1$ is suppressed for large cooperativity ${\cal C} \gg 1$. As there is no field in the cavity mode 1,  all the atoms being in state $|g_1\rangle$ is a stable solution. On the other hand,  according to the solution above with $n_{\mathrm{e}2} \approx 0$ and $n_{\mathrm{g}2}\approx 0$, mode 2 is closely resonantly excited, which leads to transmittance 1. Reversely, there is also a stable solution in which all the atoms are in $|g_2\rangle$, i.e. $n_{\mathrm{g}2} \simeq 1 $, and the transmittance of mode 1 is close to unity. The domain with 3 solutions in Fig.~\ref{fig:solnums} corresponds to the case when these stable steady states coexist. This will be further investigated along circular sections of the phase diagram, i.e., where $\eta_1^2+\eta_2^2\equiv\eta^2$ is constant. As the total input power per atom is proportional to $\eta_1^2 + \eta_2^2$, this section represents a fixed total drive intensity per atom, and increasing the angle $\varphi= \arctan{\eta_2/\eta_1}$ from 0 to $\pi/2$ corresponds to a continuous switching from driving mode 1 to 2.

Figure \ref{fig:n5000} shows cavity transmittance (a),  ground state (b) and total excited state populations (c) as a function of the angle measured from axis $\eta_1$ for case $N=5000$ along circular arcs of various radii plotted in Figure \ref{fig:solnums} (a) with the same colours. Because of the $1\leftrightarrow 2$ symmetry of the scheme, the plot of the same quantities with index 2 are just the mirror images of the ones with index 1, therefore only the latter is shown. The red and the green arcs do not cross the bistable region, hence there is only one real solution along those, which is, of course, stable. The rest of the curves, both for the transmittance and the ground state population show a characteristic S-shaped form of a bistability with overlapping stable solutions (solid line) connected by an unstable one (dashed line). High (low) transmittance corresponds to low (high) relative ground state population. For decreasing the total input power, the S-shaped curves show convergence in panels (a) and (b), whereas a gradual decrease of the population in the excited states is shown in  (c). In this limit the bistability is formed between the two hyperfine ground states, the excited states being only virtually populated underway the two-photon transition between the ground states. One can identify thus a dissipative quantum phase transition in the spirit of the three conditions given in the introduction, where in particular, the phases correspond to quantum states of high purity.

\begin{figure*}[t!!]
\raisebox{3pt}{\tikz{\node[scale=1,rectangle, inner sep=1 pt,minimum height=0.5pt,minimum width=9 pt,fill=blue](){};}} $N=5\cdot10^3$\quad
\raisebox{3pt}{\tikz{\node[scale=1,rectangle, inner sep=1 pt,minimum height=0.5pt,minimum width=9 pt,fill=orange](){};}} $N=10^4$\quad
\raisebox{3pt}{\tikz{\node[scale=1,rectangle, inner sep=1 pt,minimum height=0.5pt,minimum width=9 pt,fill=green](){};}} $N=10^5$\quad
\raisebox{3pt}{\tikz{\node[scale=1,rectangle, inner sep=1 pt,minimum height=0.5pt,minimum width=9 pt,fill=red](){};}} $N=10^6$
\centering

\includegraphics{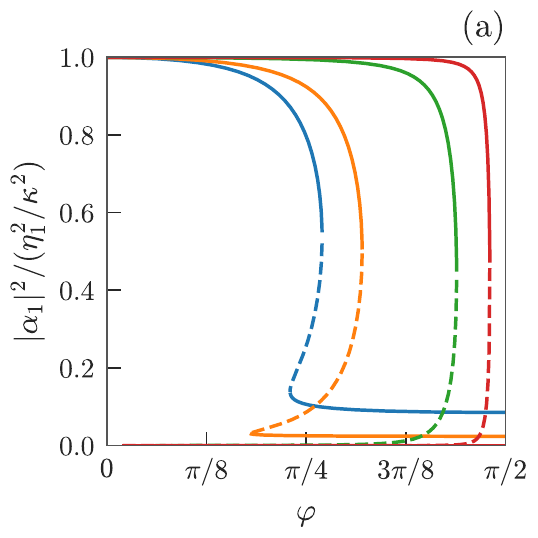}
\includegraphics{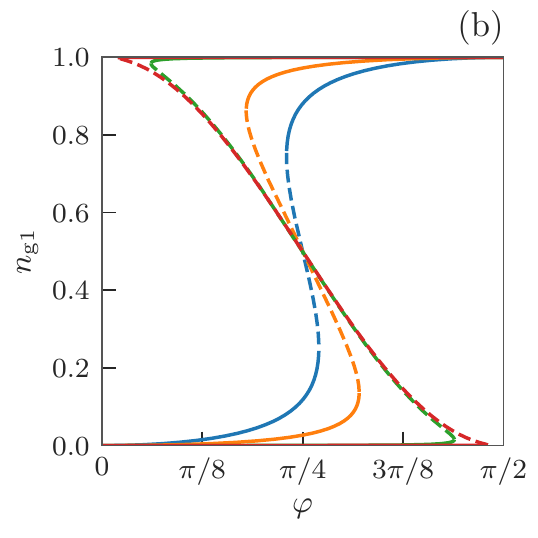}
\includegraphics{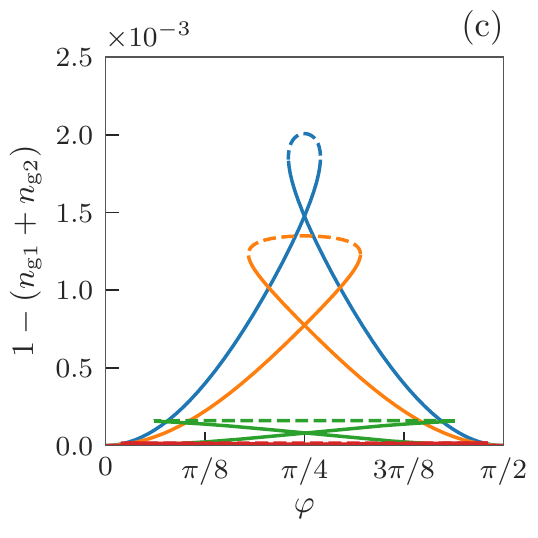}
\caption{Finite size scaling to the thermodynamic limit. The order parameter represented by the transmittance of cavity for mode 1 is plotted in (a), and relative atomic population are shown in (b-c) along the circular arcs on the phase space in Figure \ref{fig:solnums} (a) with a radius of ${\eta/\gamma=0.29}$ for different atom numbers $N$. $\varphi$ is measured from the vertical axis. Solid (dashed) lines correspond to stable (unstable) solutions.}
\label{fig:nall}
\end{figure*}

The thermodynamic limit, where the bistability becomes a phase transition can be defined as ${\cal C} \rightarrow \infty$ while $\eta$ is kept constant. In a practical case, the cooperativity can be increased by the atom number, hence the $N \rightarrow \infty$ implies   that the actual drive power $N\,\eta^2$ has to go to infinity. The axes of the phase diagrams in Figure \ref{fig:solnums} already used this scaling. Therefore the circular arcs of a radius ${\eta/\gamma=0.29}$, plotted in each phase diagram (white), are fixed in the finite-size scaling. While the boundaries of the multivalued domain vary slightly, the phase diagram is qualitatively the same. For increasing cooperativity (via atom number), Fig.~\ref{fig:nall} (a) and (b) show that the S-shaped curve tends to a sharper Z-shaped one (mirrored). Interestingly, when going towards the thermodynamic limit, both solutions become stable in almost the total range of the control parameter $\eta_2/\eta_1$. Simultaneously, as shown in Fig.~\ref{fig:nall} (c), the population in the excited states tends to completely vanish in this limit. Thus, in the thermodynamic limit, the proposed system has two stable solutions with the atoms being in one of the ground states $|g_{1(2)}\rangle$ and the other mode 2 (1) being populated by a coherent state, meaning that in this limit the perfect quantum purity of the phases of the system is achieved.


In conclusion, the proposed scheme opens the way to an experimental investigation of a first-order dissipative quantum phase transition. The four-level atomic scheme can be realized to a good approximation within the hyperfine structure of e.g. the D2 line of rubidium-87, as it has been discussed in \cite{Gabor2022}. The atom number can be varied in a controlled way over many orders of magnitude in an experiment, allowing thus for a finite-size scaling to the thermodynamic limit. Beside the possibility of investigating fundamental concepts of phase transitions in mesoscopic quantum systems, the bistability between long-lived ground states holds prospects for new atomic memory architectures.

\section*{Acknowledgements}

This research was supported by the Ministry of Culture and Innovation and the National Research, Development and Innovation Office within the Quantum Information National Laboratory of Hungary (Grant No. 2022-2.1.1-NL-2022-00004).
\bibliographystyle{unsrt}
\bibliography{TBB2}

\begin{thebibliography}{10}

\bibitem{Kessler2012}
E.~M. Kessler, G.~Giedke, A.~Imamoglu, S.~F. Yelin, M.~D. Lukin, and J.~I.
  Cirac.
\newblock Dissipative phase transition in a central spin system.
\newblock {\em Phys. Rev. A}, 86:012116, Jul 2012.

\bibitem{Macieszczak2016}
Katarzyna Macieszczak, M{\u{a}}d{\u{a}}lin Gu{\c{t}}{\u{a}}, Igor Lesanovsky,
  and Juan~P Garrahan.
\newblock Towards a theory of metastability in open quantum dynamics.
\newblock {\em Physical review letters}, 116(24):240404, 2016.

\bibitem{Lee2012}
Tony~E Lee, Hartmut Haeffner, and MC~Cross.
\newblock Collective quantum jumps of rydberg atoms.
\newblock {\em Physical review letters}, 108(2):023602, 2012.

\bibitem{Ates2012}
Cenap Ates, Beatriz Olmos, Juan~P. Garrahan, and Igor Lesanovsky.
\newblock Dynamical phases and intermittency of the dissipative quantum ising
  model.
\newblock {\em Phys. Rev. A}, 85:043620, Apr 2012.

\bibitem{Marcuzzi2014}
Matteo Marcuzzi, Emanuele Levi, Sebastian Diehl, Juan~P Garrahan, and Igor
  Lesanovsky.
\newblock Universal nonequilibrium properties of dissipative rydberg gases.
\newblock {\em Physical review letters}, 113(21):210401, 2014.

\bibitem{Carr2013}
C.~Carr, R.~Ritter, C.~G. Wade, C.~S. Adams, and K.~J. Weatherill.
\newblock Nonequilibrium phase transition in a dilute rydberg ensemble.
\newblock {\em Phys. Rev. Lett.}, 111:113901, Sep 2013.

\bibitem{Malossi2014}
N.~Malossi, M.~M. Valado, S.~Scotto, P.~Huillery, P.~Pillet, D.~Ciampini,
  E.~Arimondo, and O.~Morsch.
\newblock Full counting statistics and phase diagram of a dissipative rydberg
  gas.
\newblock {\em Phys. Rev. Lett.}, 113:023006, Jul 2014.

\bibitem{Urvoy2015}
A~Urvoy, F~Ripka, Igor Lesanovsky, D~Booth, JP~Shaffer, T~Pfau, and R~L{\"o}w.
\newblock Strongly correlated growth of rydberg aggregates in a vapor cell.
\newblock {\em Physical Review Letters}, 114(20):203002, 2015.

\bibitem{Letscher2017}
F.~Letscher, O.~Thomas, T.~Niederpr\"um, M.~Fleischhauer, and H.~Ott.
\newblock Bistability versus metastability in driven dissipative rydberg gases.
\newblock {\em Phys. Rev. X}, 7:021020, May 2017.

\bibitem{Labouvie2016}
Ralf Labouvie, Bodhaditya Santra, Simon Heun, and Herwig Ott.
\newblock Bistability in a driven-dissipative superfluid.
\newblock {\em Phys. Rev. Lett.}, 116:235302, Jun 2016.

\bibitem{Benary_2022}
J~Benary, C~Baals, E~Bernhart, J~Jiang, M~Röhrle, and H~Ott.
\newblock Experimental observation of a dissipative phase transition in a
  multi-mode many-body quantum system.
\newblock {\em New Journal of Physics}, 24(10):103034, oct 2022.

\bibitem{Ferri2021}
Francesco Ferri, Rodrigo Rosa-Medina, Fabian Finger, Nishant Dogra, Matteo
  Soriente, Oded Zilberberg, Tobias Donner, and Tilman Esslinger.
\newblock Emerging dissipative phases in a superradiant quantum gas with
  tunable decay.
\newblock {\em Phys. Rev. X}, 11:041046, Dec 2021.

\bibitem{rodriguez_probing_2017}
S.~R.~K. Rodriguez, W.~Casteels, F.~Storme, N.~Carlon~Zambon, I.~Sagnes,
  L.~Le~Gratiet, E.~Galopin, A.~Lema\^{\i}tre, A.~Amo, C.~Ciuti, and J.~Bloch.
\newblock Probing a dissipative phase transition via dynamical optical
  hysteresis.
\newblock {\em Phys. Rev. Lett.}, 118:247402, Jun 2017.

\bibitem{fink_signatures_2018}
Thomas Fink, Anne Schade, Sven Höfling, Christian Schneider, and Ataç
  Imamoglu.
\newblock Signatures of a dissipative phase transition in photon correlation
  measurements.
\newblock {\em Nature Physics}, 14(4):365--369, April 2018.

\bibitem{Dagvadorj2021}
Galbadrakh Dagvadorj, Micha\l{} Kulczykowski, Marzena~H.
  Szyma\ifmmode~\acute{n}\else \'{n}\fi{}ska, and Micha\l{} Matuszewski.
\newblock First-order dissipative phase transition in an exciton-polariton
  condensate.
\newblock {\em Phys. Rev. B}, 104:165301, Oct 2021.

\bibitem{fink_observation_2017}
J.~M. Fink, A.~Dombi, A.~Vukics, A.~Wallraff, and P.~Domokos.
\newblock Observation of the {Photon}-{Blockade} {Breakdown} {Phase}
  {Transition}.
\newblock {\em Physical Review X}, 7(1):011012, January 2017.

\bibitem{Fitzpatrick2017}
Mattias Fitzpatrick, Neereja~M. Sundaresan, Andy C.~Y. Li, Jens Koch, and
  Andrew~A. Houck.
\newblock Observation of a dissipative phase transition in a one-dimensional
  circuit qed lattice.
\newblock {\em Phys. Rev. X}, 7:011016, Feb 2017.

\bibitem{Sett2022}
Riya Sett, Farid Hassani, Duc Phan, Shabir Barzanjeh, Andras Vukics, and
  Johannes~M Fink.
\newblock Emergent macroscopic bistability induced by a single superconducting
  qubit.
\newblock {\em arXiv preprint arXiv:2210.14182}, 2022.

\bibitem{Bonifacio1978photon}
R~Bonifacio and LA~Lugiato.
\newblock Photon statistics and spectrum of transmitted light in optical
  bistability.
\newblock {\em Physical Review Letters}, 40(15):1023, 1978.

\bibitem{Bonifacio1978bistable}
R.~Bonifacio and L.~A. Lugiato.
\newblock Bistable absorption in a ring cavity.
\newblock {\em Lettere al Nuovo Cimento (1971-1985)}, 21(15):505--509, Apr
  1978.

\bibitem{dombi_optical_2013}
András Dombi, András Vukics, and Peter Domokos.
\newblock Optical bistability in strong-coupling cavity {QED} with a few atoms.
\newblock {\em Journal of Physics B: Atomic, Molecular and Optical Physics},
  46(22):224010, 2013.
\newblock Publisher: IOP Publishing.

\bibitem{Casteels2017}
W.~Casteels, R.~Fazio, and C.~Ciuti.
\newblock Critical dynamical properties of a first-order dissipative phase
  transition.
\newblock {\em Phys. Rev. A}, 95:012128, Jan 2017.

\bibitem{Heugel2019}
Toni~L. Heugel, Matteo Biondi, Oded Zilberberg, and R.~Chitra.
\newblock Quantum transducer using a parametric driven-dissipative phase
  transition.
\newblock {\em Phys. Rev. Lett.}, 123:173601, Oct 2019.

\bibitem{dombi_bistability_2015}
{Dombi, András}, {Vukics, András}, and {Domokos, Peter}.
\newblock Bistability effect in the extreme strong coupling regime of the
  {Jaynes}-{Cummings} model.
\newblock {\em Eur. Phys. J. D}, 69(3):60, 2015.

\bibitem{carmichael_breakdown_2015}
H.~J. Carmichael.
\newblock Breakdown of {Photon} {Blockade}: {A} {Dissipative} {Quantum} {Phase}
  {Transition} in {Zero} {Dimensions}.
\newblock {\em Phys. Rev. X}, 5(3):031028, September 2015.
\newblock Publisher: American Physical Society.

\bibitem{vukics_finite_2019}
A.~Vukics, A.~Dombi, J.~M. Fink, and P.~Domokos.
\newblock Finite-size scaling of the photon-blockade breakdown dissipative
  quantum phase transition.
\newblock {\em {Quantum}}, 3:150, June 2019.

\bibitem{Clark2022}
T.~W. Clark, A.~Dombi, F.~I.~B. Williams, \'A. Kurk\'o, J.~Fort\'agh, D.~Nagy,
  A.~Vukics, and P.~Domokos.
\newblock Time-resolved observation of a dynamical phase transition with atoms
  in a cavity.
\newblock {\em Phys. Rev. A}, 105:063712, Jun 2022.

\bibitem{Gabor2022}
B.~G\'abor, D.~Nagy, A.~Dombi, T.~W. Clark, F.~I.~B. Williams, K.~V. Adwaith,
  A.~Vukics, and P.~Domokos.
\newblock Ground-state bistability of cold atoms in a cavity.
\newblock {\em Phys. Rev. A}, 107:023713, Feb 2023.

\bibitem{Suarez2023collective}
Elmer Suarez, Federico Carollo, Igor Lesanovsky, Beatriz Olmos, Philippe~W
  Courteille, and Sebastian Slama.
\newblock Collective atom-cavity coupling and nonlinear dynamics with atoms
  with multilevel ground states.
\newblock {\em Physical Review A}, 107(2):023714, 2023.

\end{thebibliography}

\end{document}